\documentclass[11pt]{article}

\usepackage{amssymb, amsmath,fullpage} 
\usepackage{graphicx}

\usepackage{url}
\newtheorem{Remark}{Remark} 
\newtheorem{Example}{Example} 
\newtheorem{thm}{Theorem} 
\newtheorem{dfn}{Definition} 
\newtheorem{prp}{Proposition}

\newenvironment{proof}{\paragraph{Proof:}}{\hfill$\square$}

\usepackage{url}

\title{Scales of Fr\'echet means and Karcher quasi-arithmetic means}

\author{Frank \textsc{Nielsen}\footnote{Sony Computer Science Laboratories Inc, Tokyo 141-0022, Japan.\newline e-mail: \texttt{Frank.Nielsen@acm.org}}
          }
 \date{}

\def\dmu{\mathrm{d}\mu}
\def\dt{\mathrm{d}t}

\def\bbR{\mathbb{R}}

\def\inner#1#2{\langle #1,#2\rangle}

\def\st{\ :\ }

\def\bbR{\mathbb{R}}
\def\ds{\mathrm{d}s}

\def\dlambda{\mathrm{d}\lambda}
\def\dtheta{\mathrm{d}\theta}
\def\Euc{\mathrm{Euc}}

\def\st{\ :\ }
\def\du{{\mathrm{d}u}}
\def\sqr{\mathrm{sqr}}
\def\dtheta{\mathrm{d}\theta}
\def\deta{\mathrm{d}\eta}

\begin{document}
%

\maketitle

\begin{abstract}      
In this paper, we first prove that any interior point of an open interval of the real line can be interpreted as Fr\'echet  means with respect to corresponding metric distances, thus extending the result of [Dinh et al., Mathematical Intelligencer 47.2 (2025)] which was restricted  to intervals on the positive reals by
 using the family of power means: 
Our generic construction relies on the concept of scales of means that we demonstrate with the scale of exponential means and the scale of radical means.
Second, we interpret those Fr\'echet means geometrically as the center of mass of any two distinct points on the Euclidean line expressed in various coordinate systems: 
Namely, by interpreting the Euclidean line as a 1D Hessian Riemannian manifold, we introduce pairs of dual Fr\'echet/Karcher means related by convex duality in dual coordinate systems.
This result yields us to consider squared Hessian metrics in arbitrary dimension:
 We prove that these squared Hessian metrics amount to Euclidean geometry with the Riemannian center of mass expressed in primal coordinate systems as multivariate quasi-arithmetic means coinciding with left-sided Bregman centroids.
\end{abstract}


\section{Introduction: Distances and Fr\'echet means} 

Din, Tran, and Truong~\cite{dinh2025every} recently showed that every interior point $c$ of a finite interval $[a,b]$ in $[0,\infty)$ can be interpreted as the midpoint with respect to some metric distance, where

\begin{dfn}
A dissimilarity measure $d(x,y)$ is mathematically called a {\em distance} if and only if it satisfies the four  metric axioms:
(i) Non-negativity: $d(x,y)\geq 0$, 
(ii) Identity of the indiscernibles: $d(x,y)=0$ if and only if $x=y$,
	(iii) Symmetry $d(x,y)=d(y,x)$, 
	and (iv) triangle inequality: $d(x,z)+d(z,y)\geq d(x,y)$ for all $z$
	\end{dfn}
	
\begin{dfn} 
A point $c\in (a,b)$ is said to be a {\em midpoint} of $a$ and $b$ with respect to a distance $d(\cdot,\cdot)$ if $d(a,c)=d(c,b)$. 
	\end{dfn}
	
More precisely, Din, Tran, and Truong~\cite{dinh2025every} proved that any $c\in(a,b)\subset [0,+\infty)$  
	is the midpoint with respect to the distance $d_{p}(x,y)=|x^p-y^p|$.
That is, for a given $c$, there exists a power exponent $p$ depending on $c$ (i.e., $p=p(c)$ which uniquely exists but has no closed-form expression) such that
  $c=\left(\frac{a^p+b^p}{2}\right)^{\frac{1}{p}}$ is the power mean midpoint 
	satisfying $d_p(a,c)=d_p(c,b)$ (Theorem~1 of~\cite{dinh2025every}).

In this work, we first give a {\em generalization} of their theorem using the 
concept of {\em scales of means}~\cite{pasteczka2012family} which allows us to consider any interval on the real line $\bbR$: 
Our generic Theorem~\ref{thm:scale} is then instantiated with the scale of {\em exponential means} in Theorem~\ref{thm:scale-em} and with the scale of  {\em radical means} in Theorem~\ref{thm:scale-rm}. 
Second, we interpret 
those midpoints as the various representations in corresponding coordinate systems of the same Euclidean center of mass of two prescribed distinct points lying on the Euclidean line in~\S\ref{sec:geo} (Proposition~\ref{prop:qam}).
This interpretation relies on viewing the Euclidean line $\bbR$ as a 1D Hessian Riemannian manifold which let us highlight the novel notion of {\em dual scales of means} arising from convex duality of potential functions in \S\ref{sec:dualscales} (Proposition~\ref{dualscales}).
Furthermore, we consider squared Hessian metrics in \S\ref{sec:KarcherQAM} and show that the geodesic distance amounts to the Euclidean distance expressed in dual coordinate systems in Proposition~\ref{prp:sqrHessianeucldist}, and that the center of mass (Karcher mean when considered as a Riemannian center of mass) can be expressed as a multivariate quasi-arithmetic mean in the primal coordinate system (Proposition~\ref{prp:mvqam}).
Finally, we conclude with a discussion from the viewpoint of information geometry in \S\ref{sec:IG}.

Let us start with the generic definition of a mean~\cite{kolmogorov1930notion,de2016mean}:

\begin{dfn}\label{def:mean}
A mean $m(x_1,\ldots,x_n)$ is a $n$-variate function $m: I^n\rightarrow I$ with domain $I^n\subset\bbR^n$ that satisfies the following properties: 
(i)  $m(x,\ldots,x)=x$ (idempotence), (ii)  $\min\{x_1,\ldots,x_n\} \leq m(x_1,\ldots,x_n)\leq \max\{x_1,\ldots,x_n\}$ (internality), and 
(iii)  $m(x_1,\ldots,x_n)=m(\sigma(x_1,\ldots,x_n))$ for any permutation $\sigma$ (symmetry).
\end{dfn}

In the reminder, we consider additive metric distances satisfying $d(a,x)+d(x,b)=d(a,b)$ for all $x\in(a,b)$.
In that case, the midpoint~\cite{dinh2025every} is the unique Fr\'echet mean~\cite{Frechet-1948}:

\begin{dfn}
The Fr\'echet mean(s) $c$ of two points $a$ and $b$ with respect to a distance $d(\cdot,\cdot)$ is defined by:
$$
c=\arg\min_{x\in [a,b]} d^2(a,x)+d^2(x,b).
$$
\end{dfn}

Note that in general, the Fr\'echet mean in a metric space $(X,d)$ may not be unique. 
For example, two antipodal points on a 3D sphere have a great circle as Fr\'echet means with respect to the sphere geodesic metric distance.

Instead of using the power means to realize the midpoints $c$ which constrains intervals $(a,b)$ to be on the positive reals, we shall consider broader families of means called quasi-arithmetic means~\cite{bullen2003quasi}:

\begin{dfn}
A {\em quasi-arithmetic mean} $m_h(a,b)$ is defined according to any continuous strictly monotone scalar function $h$ by  $m_h(a,b)=h^{-1}\left(\frac{h(a)+h(b)}{2}\right)$. 
\end{dfn}

In general, the quasi-arithmetic mean can be extended to $n$ variables as 
$$
m_h(x_1,\ldots,x_n)=h^{-1}\left(\frac{1}{n}\sum_i h(x_i)\right).
$$
Note that if a mean $m(x_1,\ldots,x_n)$ (Definition~\ref{def:mean}) is continuous and increasing in each variable and further satisfies the absorbing property that
$$
m(\underbrace{s,\ldots,s}_{k},x_{k+1},\ldots,x_n)=s,
$$ 
for any $k\in\{2,\ldots,n\}$ and $s=m(x_1,\ldots,x_n)$ then the mean is necessarily a quasi-arithmetic mean~\cite{de2016mean}.

We have $m_h=m_l$ if and only if there exists constants $\alpha\not=0$ and $\beta\in\bbR$ such that $h(u)=\alpha l(u)+\beta$.
In particular, one can check that $m_h=m_{-h}$.
Quasi-arithmetic means are regular means: 
They satisfy (i) the internality property (i.e., $\min\{a,b\}\leq m_h(a,b)\leq \max\{a,b\}$),
 (ii) the strictness property (i.e., equals an input only if all inputs are equal),
(iii) the 
continuity property (i.e., no jumps $(a',b')\rightarrow (a,b) \Rightarrow m_h(a',b')\rightarrow m_h(a,b)$),
(iv) the symmetry property (i.e., $m_h(a,b)=m_h(b,a)$),
and (v) the monotonicity property (i.e., if $a'\leq a$ and $b'\leq b$ then $m_h(a',b')\leq m_h(a,b)$).

Power means are quasi-arithmetic means obtained by the following corresponding family of generators:
$$
m_{h_p}(x,y)=\left\{
\begin{array}{ll}
\left(\frac{x^p+y^p}{2}\right)^{\frac{1}{p}}, & p\not=0,\\
\sqrt{xy}, & p=0.
\end{array}
\right.,
$$
where
$$
h_p(u)=\left\{
\begin{array}{ll}
u^p, & p\not=0,\\
\log u, & p=0.
\end{array}
\right.
$$

The Fr\'echet mean with respect to the 1D distance $d_f(x,y)=|f(x)-f(y)|$  for a positive differentiable strictly monotone function $f$ on $\bbR_{\geq 0}$ is (Lemma~1 of~\cite{dinh2025every}):
\begin{equation}\label{eq:qam}
c_{d_f}(a,b) = f^{-1}\left(\frac{f(a)+f(b)}{2}\right).
\end{equation}

That is, the midpoint $c_{d_f}(a,b)$ with respect to distance $d_f$ is a quasi-arithmetic mean~\cite{bullen2003quasi}:  $c_{d_f}(a,b)=m_{f}(a,b)$.
In particular, the power means $m_{h_p}$ are midpoints with respect to the distances 
$$
d_{h_p}(x,y)=
\left\{
\begin{array}{ll}
|x^p-y^p|, & p\not =0,\\
|\log x-\log y|, & p=0.
\end{array}
\right.
$$

\section{Midpoints from scales of means}

Instead of considering the power mean construction of~\cite{dinh2025every} which limits $(a,b)$ on the positive reals, we may use any arbitrary scale of means~\cite{persson1990generalized}:

\begin{dfn} 
A {\em scale of means} is a one-parameter family of means $\{m_r(\cdot,\cdot)\}_{r\in\bbR}$ with $m_r:I^2\subset\bbR^2\rightarrow I$ such that for all $a,b\in I$, we have:
(i) $r\mapsto m_r(a,b)$ is continuous on $\bbR$,
(ii) $r\mapsto m_r$ is strictly monotone, and 
(iii)  either $\lim_{r\rightarrow -\infty} m_r(a,b)=\min\{a,b\}$ and $\lim_{r\rightarrow +\infty} m_r(a,b)=\max\{a,b\}$ (increasing scale), or
$\lim_{r\rightarrow-\infty} m_r(a,b)=\max\{a,b\}$ and $\lim_{r\rightarrow +\infty} m_r(a,b)=\min\{a,b\}$ (decreasing scale).
\end{dfn}

The family of power means $\{m_p\}_{p\in\bbR}$ with $m_p: \bbR_{>0}\rightarrow \bbR$ are the only  {\em positively homogeneous} quasi-arithmetic means 
 defined on the positive real domains $I=\bbR_{>0}$ (i.e., $m_p(\lambda a,\lambda b)=\lambda m_p(a,b)$ for any $\lambda>0$) which forms an increasing scale (see proof in~\cite{pasteczka2015scales}).
Thus we can solve $m_p(a,b)=c$ equivalently as $m_p(1,\frac{b}{a})=\frac{c}{a}$: Although there is no closed-form solution, we can numerically approximate the unique solution, say,  using the Newton-Raphson method.

Since the power means form an increasing scale, we get the QM-AM-GM-HM inequalities between the harmonic mean (HM), geometric mean (GM), arithmetic mean (AM), and quadratic mean (QM): 
$$
\mathrm{QM}\geq \mathrm{AM}\geq \mathrm{GM}\geq \mathrm{HM}.
$$

Let us state our theorem which generalizes and extends Theorem~1 of~\cite{dinh2025every}:

\begin{thm}\label{thm:scale}
Let $\{s_\alpha\}_{\alpha\in\bbR}$ be a family of strictly monotone and differentiable functions yielding a scale $\{m_{s_\alpha}\}$ of quasi-arithmetic means, where $m_{s_\alpha}: I\times I\rightarrow I$ for $I\subset\bbR$.
Then for any scalar $c$ of a given interval $(a,b)\subset I$ there exists a corresponding parameter $\alpha=\alpha(a,b,c)$ (unique when the scale is stricty monotone with respect to $\alpha$) such that
$c=m_{s_\alpha}(a,b)=s_\alpha^{-1}\left(\frac{s_\alpha(a)+s_\alpha(b)}{2}\right)$.
Therefore $c$ is the midpoint of $a$ and $b$ with respect to distance $d_{s_\alpha}(x,y)=|s_\alpha(x)-s_\alpha(y)|$: $d_{s_\alpha}(a,c)=d_{s_\alpha}(c,b)$.
\end{thm}

Notice that there are many non quasi-arithmetic means which form scale of means~\cite{bullen2013handbook} like the Lehmer means, the Stolarsky means, the identric means, etc.

We now instantiate Theorem~\ref{thm:scale} to an increasing scale and a decreasing scale of quasi-arithmetic means.

\subsection{Example 1: Increasing scale of exponential means on $I=\bbR$}

Using a different scale of means than the scale of power means allows one to prove that for any $(a,b)\in\bbR$, there exists a family of distances $d_{e_\alpha}$ with corresponding midpoints covering the interval $(a,b)\subset\bbR$.
Let us consider the family of quasi-arithmetic exponential means induced by the generators:

$$
{e_\alpha}(u)=\left\{
\begin{array}{ll}
e^{\alpha u}, & \alpha\not=0,\\
{u}, & \alpha=0
\end{array}
\right.
$$
for $\alpha\in\bbR$ with corresponding inverse functions:
$$
{e_\alpha}^{-1}(u)=\left\{
\begin{array}{ll}
\frac{1}{\alpha}\log u, & \alpha\not=0,\\
{u}, & \alpha=0
\end{array}
\right.
$$

We get the  family $\{m_{e_\alpha}\}_{\alpha\in\bbR}$ of quasi-arithmetic {\em exponential means}:
\begin{equation}
m_{e_\alpha}(x,y)=\left\{
\begin{array}{ll}
\frac{1}{\alpha}\log\left(\frac{e^{\alpha x}+e^{\alpha y}}{2}\right), & \alpha\not= 0\\
\frac{x+y}{2}, & \alpha=0.
\end{array}
\right.
\end{equation}
This family forms an increasing scale in $\bbR$ (see Remark~2 of~\cite{pasteczka2012family}.

Notice that the exponential mean $m_{e_\alpha}(x,y)$ is a scaled log-sum-exp (LSE) biparametric function such that
when $\alpha$ is large enough, we have $\frac{e^{\alpha x}+e^{\alpha y}}{2} \approx \frac{e^{\alpha\max\{x,y\}}}{2}$ and thus 
$m_{e_\alpha}(x,y)\approx \frac{1}{\alpha}(\log e^{\alpha\max\{x,y\}}-\log 2)\approx \max\{x,y\}$ with $\lim_{\alpha\rightarrow\infty} m_{e_\alpha}(x,y)=\max\{x,y\}$.
That is, for large enough $\alpha$, $m_{e_\alpha}(x,y)$ is a differentiable approximation of the non-differentiable maximum bivariate function.
Similarly, when $\alpha$ is tending toward $-\infty$, we have $m_{e_\alpha}(x,y)\approx \min\{x,y\}$ and in the limit case, we get 
$\lim_{\alpha\rightarrow-\infty} m_{e_\alpha}(x,y)=\min\{x,y\}$.

The distances corresponding  to $\{m_{e_\alpha}(x,y)\}_\alpha$ are given by
\begin{equation}
d_{e_\alpha}(x,y)=
\left\{
\begin{array}{ll}
|e^{\alpha x}-e^{\alpha y}|, & \alpha\not=0,\\
|x-y|, & \alpha=0.
\end{array}
\right.
\end{equation}

Thus we get the following instance of Theorem~\ref{thm:scale}:

\begin{thm}\label{thm:scale-em}
The midpoints $\{c_{e_\alpha}(a,b)\}_{\alpha\in\bbR}$ with respect to distance $d_{e_\alpha}$ range in $(a,b)$ for any $-\infty<a<b<\infty$.
\end{thm}

\subsection{Example 2: Decreasing scale of radical means on $\bbR_{>0}$}

Consider the family of quasi-arithmetic  radical means~\cite{pasteczka2012family} $\{m_{k_\alpha}\st \alpha\in\bbR\}$ induced by the generators:
$$
{k_\alpha}(u)=\left\{
\begin{array}{ll}
\alpha^{\frac{1}{u}}=\exp(\frac{1}{u}\log\alpha), & \alpha>0, \alpha\not=1,\\
\frac{1}{u}, & \alpha=1
\end{array}
\right.
$$
for $\alpha\in\bbR_{>0}$ with reciprocal functions:
$$
{k_\alpha}^{-1}(u)=\left\{
\begin{array}{ll}
\frac{\log\alpha}{\log u}, & \alpha\not=1,\\
\frac{1}{u}, & \alpha=1
\end{array}
\right.
$$

That family forms a decreasing scale of means on $I=\bbR_{>0}$ with
\begin{equation}
m_{k_\alpha}(x,y)=\left\{
\begin{array}{ll}
 \left( \frac{1}{\log\alpha} \, \log\left( \frac{ \alpha^{\frac{1}{a}} + \alpha^{\frac{1}{b}} }{2} \right) \right)^{-1}, & \alpha\not= 1\\
\frac{2xy}{x+y}, & \alpha=1.
\end{array}
\right.
\end{equation}

In particular, mean $m_{k_1}$ is the harmonic mean (HM).

The induced radical metric distances are given by
\begin{equation}
d_{k_\alpha}(x,y)=|k_\alpha(x)-k_\alpha(y)|=\left\{
\begin{array}{ll}
|\alpha^{\frac{1}{x}}-\alpha^{\frac{1}{y}}|, & \alpha\not=1,\\
\left|\frac{1}{x}-\frac{1}{y}\right|, & \alpha=1.
\end{array}
\right.
\end{equation}

Notice that we can reparameterize the radical scale using the transformation $\beta=\log\alpha\in\bbR$ (with $\alpha=e^\beta$).
It follows the family of generators
$$
r_\beta(u)=\left\{
\begin{array}{ll}
\ \exp(\frac{\beta}{u}), & \beta\not=0,\\
\frac{1}{u}, & \beta=0
\end{array}
\right.
$$
for $\beta\in\bbR$ with reciprocal functions:
$$
{r_\beta}^{-1}(u)=\left\{
\begin{array}{ll}
\frac{\beta}{\log u}, & \beta\not=0,\\
\frac{1}{u}, & \beta=0
\end{array}
\right.,
$$
and the radical mean
$$
m_{r_\beta}(x,y)=\left\{
\begin{array}{ll}
 \frac{\beta}{\log\left( \frac{ e^{\frac{\beta}{a} } + e^{\frac{\beta}{b} } }{2} \right)}, & \beta\not= 0\\
\frac{2xy}{x+y}, & \beta=0.
\end{array}
\right.
$$

\begin{figure}%
\centering
\includegraphics[width=0.7\columnwidth]{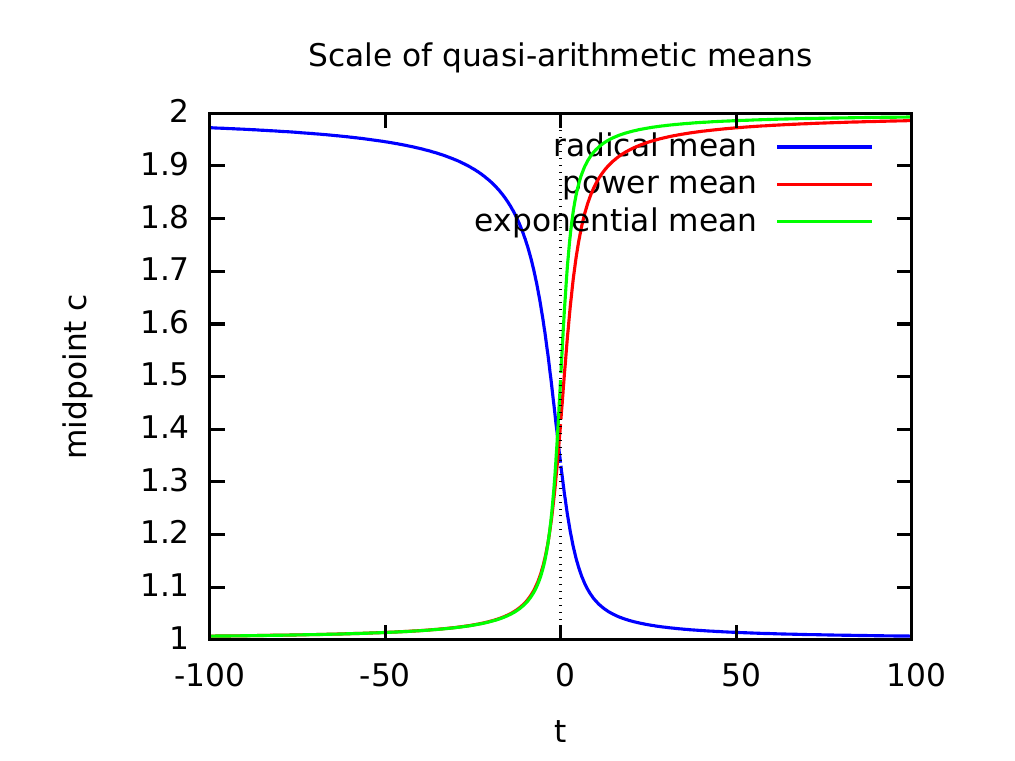}%

\caption{Plots of scales of means for the power, exponential, and radical means: (left) scale parameter $t$ (horizontal axis) ranges in $[-100,100]$ showing different stretching properties of the scales covering the interval $(a=1,b=2)$ on the vertical axis.}
\label{fig:scales}%
\end{figure}

Figure~\ref{fig:scales} plots the  scales for the power, exponential, and radical family of means (i.e., $\{m_{h_t}\}_t$, $\{m_{e_t}\}_t$, and $\{m_{r_t}\}_t$, respectively for $t\in\bbR$) for prescribed values of $a=1$ and $b=2$.

Thus we have shown a different realization of the power mean result of~\cite{dinh2025every} for a family of metric distances $\{d_{k_\alpha}\}_{\alpha\in\bbR}$ with midpoints covering the range $(a,b)\subset\bbR_{>0}$.

We summarize the result by the following instance of Theorem~\ref{thm:scale}:

\begin{thm}\label{thm:scale-rm}
The midpoints $\{c_{k_\alpha}(a,b)\}_{\alpha\in\bbR_{>0}}$ with respect to distance $d_{k_\alpha}$ for $\alpha\in\bbR_{>0}$ range in $(a,b)$ for any $0<a<b<\infty$.
\end{thm}

\section{A geometric Riemannian interpretation}\label{sec:geo} 

Those quasi-arithmetic mean midpoints can be interpreted to correspond to the same Euclidean center of mass 
$C=\frac{A+B}{2}$ of two prescribed distinct points $A$ and $B$ expressed in various coordinate systems:

Consider a coordinate system $(I,x)$ such that $x(A)=a$ and $x(B)=b$ with $a<b$.
Let $(I',x')$ be another coordinate system such that $x=h(x')$ for a strictly monotone and continuous function $h$ with $x'(x)=h^{-1}(x)$.
The center of mass is expressed in the $x$-coordinate system as $x(C)=\frac{x(A)+x(B)}{2}=c$, i.e., $c=\frac{a+b}{2}$, the arithmetic mean.
Since $c=h(c')$, $a=h(a')$, and $b=h(b')$ where $a'$, $b'$, and $c'$ are the coordinates of $A$, $B$, and $C$ in the $x'$-coordinate system, 
we have $h(c')=\frac{h(a')+h(b')}{2}$, i.e., $c'=m_h(a',b')$ since $h$ is an invertible diffeomorphism.
Therefore $c'$ corresponds to the $x'$-coordinate of the Euclidean center of mass $C$ in the chart $(I',x'(\cdot))$.
Thus the scale $\{m_{s_\alpha}\}_{\alpha}$ of quasi-arithmetic means represent the same Euclidean center of mass $C$ of two points $A$ and $B$ on the Euclidean line in a corresponding family of charts $(I_\alpha,x_\alpha(\cdot))$.

We shall now consider the Euclidean line as a  1D Riemannian manifold equipped with a Hessian metric~\cite{amari2014curvature} to derive dual scales of means.

\section{Dual scales of means from convex analysis}\label{sec:dualscales}

Last, let us reconsider the 1D Euclidean line from the viewpoint of Riemannian geometry:
Let $(M,g)$ be a 1D Riemannian manifold with the Riemannian metric $g$ expressed in the global coordinate system $(\bbR,\theta)$ by $g_{11}(\theta)>0$.
Then $g$ is a Hessian metric~\cite{Shima-2007}, i.e., there exists a strictly convex and smooth potential function $f(\theta)$ such that $g(\theta)=f''(\theta)>0$. 
It follows that the length element $\ds$ is $\sqrt{g_{11}(\theta)}\,\dtheta$, and the Riemannian geodesic metric distance $\rho(\cdot,\cdot)$ is given by:
$$
\rho(\theta_1,\theta_2) = \int_{\theta_1}^{\theta_2} \sqrt{g_{11}(u)}\, \du.
$$

Let $h(\theta)=\int^\theta\sqrt{g_{11}(u)}\du =\int^\theta \sqrt{f''(u)}\du$ be an antiderivative of $\sqrt{f''}$.
Function $h$ is a strictly increasing function since $h'(\theta)=\sqrt{f''(\theta)}>0$.
Thus we have the 1D Riemannian distance expressed as follows:

\begin{equation}
\rho(\theta_1,\theta_2)= \left| h(\theta_2)-h(\theta_1) \right|.
\end{equation}

\begin{prp}\label{prop:qam}
Let $(M,g)$ be a 1D Riemannian manifold with Hessian metric expressed in the $\theta$-coordinate system as $g(\theta)=f''(\theta)>0$.
Then the Riemannian center of mass $C$  of two points $A$ and $B$ with coordinates $\theta(A)=a$ and $\theta(B)=b$ is
\begin{equation}\label{eq:hqam}
\theta(C)=h^{-1}\left( \frac{ h(a) + h(b)}{2}\right)=m_h(a,b).
\end{equation}
\end{prp}

\begin{proof}
Consider the Riemannian center of mass  or Karcher mean~\cite{karcher2014riemannian} of two points $A$ and $B$ on $(M,g)$ with coordinates $\theta(A)=\theta_1=a$ and $\theta(B)=\theta_2=b$, respectively.
The Riemannian center of mass is the least squares minimizer of the sum (or equivalently average) squared Riemannian distances:

$$
 \rho^2(\theta_1,\theta) + \rho^2(\theta_2,\theta) 
=  (h(\theta_1)-h(\theta))^2 +  (h(\theta_2)-h(\theta))^2.
$$

This minimization problem is equivalent to minimize the following energy function:
$$
E(\theta)=-2 h(\theta)\, \left(h(\theta_1)+h(\theta_2)\right) + 2\,h^2(\theta).
$$

Setting the derivative of $E(\theta)$ to zero, we get:
$$
-2h'(\theta)\,\left(h(\theta_1)+h(\theta_2)\right)+4\,h(\theta)h'(\theta)=0.
$$
Since $h'(\theta)>0$, we obtain $h(\theta)=\left(\frac{h(\theta_1)+h(\theta_2)}{2}\right)$.

Hence, we find that the Riemannian centroid of $A$ and $B$ (called the Karcher mean~\cite{karcher2014riemannian}) is expressed in the $\theta$-coordinate system by a quasi-arithmetic mean:

\begin{equation} 
\theta=h^{-1}\left( \frac{ h(\theta_1) + h(\theta_2)}{2}\right)=m_h(a,b).
\end{equation}
\end{proof}

Now, the metric $g(\theta)=g_{11}(\theta)$ is in fact the Euclidean metric (written in the Cartesian coordinate system $(\bbR,\lambda)$ as $g_\Euc(\lambda)=1$) since 
 the following metric change of coordinate transformation holds:
$$
g(\theta)=  g_\Euc(\lambda(\theta)) \, \left(\frac{\dlambda}{\dtheta} \right)^2 = \left(\frac{\dlambda}{\dtheta} \right)^2.
$$
It follows that $\frac{\dlambda}{\dtheta}=\sqrt{g(\theta)}$ and we recover $\lambda(\theta)=\int^\theta\sqrt{g(u)}\du=h(\theta)$.

Consider now the Legendre convex conjugate~\cite{Shima-2007} $f^*(\eta)$ of $f(\theta)$
$$
f^*(\eta) = \theta(\eta)\eta-f(\theta(\eta)),
$$
such that $\eta(\theta)=f'(\theta)$ and $\theta(\eta)={f^*}'(\eta)$.
We have the Euclidean metric $g_\Euc(\lambda)=1$ which can be expressed in these dual $\theta/\eta$-coordinate systems as 
$g(\theta)=f''(\theta)$ and $g^*(\eta)={f^*}''(\eta)$ (with the Crouzeix identities~\cite{crouzeix1977relationship}: $g(\theta) g^*(\eta(\theta))= g(\theta(\eta)) g^*(\eta)= 1$).

The Euclidean center of mass $C$ expressed in the $\theta$-coordinate system is $m_h(a,b)$ with $h=\int^\theta \sqrt{f''(u)}\,\du$.
It can be expressed in the dual $\eta$-coordinate system as $m_{h^\diamond}(a',b')$ where $a'=f'(a)$ and $b'=f'(b)$ and 
$h^\diamond(u)=\int^\eta \sqrt{{f^*}''(u)}\,\du$.

\begin{prp}[Dual quasi-arithmetic means]\label{dualscales}
For a Legendre-type scalar function~\cite{LegendreType-1967} $f$ with convex conjugate $f^*$, the following dual quasi-arithmetic mean identities hold:
\begin{equation}\label{eq:dualqam}
f'(m_h(a,b))=m_{h^\diamond}(f'(a),f'(b)), \quad m_h(a,b)={f^*}'\left(m_{h^\diamond}(f'(a),f'(b))\right),
\end{equation}
where $h=\int^\theta \sqrt{f''(u)}\,\du$ and $h^\diamond(u)=\int^\eta \sqrt{{f^*}''(u)}\,\du$.
\end{prp}

Thus when considering a scale of means $\{m_{s_\alpha}\}_\alpha$, we can consider equivalently its dual scale of means $\{m_{s_\alpha^\diamond}\}_\alpha$ on the dual parameters.

Let us  illustrate this result with two examples of  pairs of quasi-arithmetic means linked by convex duality:

\begin{Example}\label{ex1}
Consider $f(\theta)=e^\theta$ with $\eta(\theta)=f'(\theta)=e^\theta$ and $g(\theta)=f''(\theta)=e^\theta$.
The convex conjugate is $f^*(\eta)=\eta\log\eta-\eta$ (negative Shannon entropy) with ${f^*}'(\eta)=\log\eta$ and $g^*(\eta)={f^*}''(\eta)=\frac{1}{\eta}$.
We have $h(\theta)=\int^\theta\sqrt{f''(u)}\,\du=2\exp(\theta/2)$ with $h^{-1}(\theta)=2\log\frac{\theta}{2}$.
Similarly, we have $h^\diamond(\eta)=\int^\eta\sqrt{{f^*}''(u)}\,\du=2\sqrt{\eta}$ with ${h^\diamond}^{-1}(\eta)=(\frac{\eta}{u})^2$.
We get the following pair of quasi-arithmetic means satisfying Eq.~\ref{eq:dualqam}:

\begin{eqnarray*}
m_h(a,b) &=& 2\log\frac{e^{a/2}+e^{b/2}}{2},\\
m_{h^\diamond}(a',b') &=& \left(\frac{\sqrt{a'}+\sqrt{b'}}{2}\right)^2.
\end{eqnarray*}

We check that this pair of quasi-arithmetic means are in duality as follows:
$$
f'(m_h(\theta_1,\theta_2)) = \exp\left( 2\, \log\frac{e^{\theta_1/2}+e^{\theta_2/2}}{2} \right) = m_{h^\diamond}(\eta_1,\eta_2),
$$
where $\eta_i=f'(\theta_i)$ and $\theta_i={f^*}'(\eta_i)$.
\end{Example}

\begin{Example}\label{ex2}
Let $f(\theta)=\log(1+e^\theta)$ with $f'(\theta)=\frac{e^\theta}{1+e^\theta}$ and $f''(\theta)=\frac{e^\theta}{(1+e^\theta)^2}$.
We get $h(\theta)=\int^\theta \sqrt{f''(u)}\,\du= 2\mathrm{arctan}(\exp(\theta/2))$ and 
$h^{-1}(\theta)=2\log(\tan(\theta/2))$.
The convex conjugate is $f^*(\eta)=\eta\log\eta+(1-\eta)\log(1-\eta)$ with ${f^*}'(\eta)=\log\frac{\eta}{1-\eta}$ and ${f^*}''(\eta)=\frac{1}{\eta(1-\eta)}$.
It follows that $h^\diamond(\eta)=\int^\eta \sqrt{{f^*}''(u)}\,\du=2\,\mathrm{arcsin}(\sqrt{\eta})$ with ${h^\diamond}^{-1}(\eta)=\sin^2\left(\frac{\eta}{2}\right)$.
We check that $f'(m_h(\theta_1,\theta_2))=m_{h^\diamond}(\eta_1,\eta_2)$ (Eq.~\ref{eq:dualqam}).
\end{Example}

\begin{Remark}
The scalar quasi-arithmetic mean $m_h(a,b)$ can also be interpreted as the left-sided Bregman centroid~\cite{nielsen2009sided}
$$
m_{\phi'}(a,b)=\arg \min_c B_\phi(c,a)+B_\phi(c,b),
$$
with respect to the Bregman divergence~\cite{Bregman-1967}: 
$$
B_\phi(x,y)=\phi(x)-\phi(y)-(x-y)\phi'(y),
$$ 
for the strictly convex and differentiable generator $\phi(u)=\int^u h(u) \du$ (with $h=\phi'$ so that  $m_{h}(a,b)=m_{\phi'}(a,b)$). 
Notice that Hessian manifolds have canonical divergences which can be expressed as Bregman divergences~\cite{IG-2016,nielsen2022many}.
In case of separable $m$-dimensional Bregman divergences $\phi(\theta)=\sum_{i=1}^m \phi_i(\theta_i)$, we have $c_i=m_{h_i}(a_i,b_i)$where $h_i=\phi_i'$.
\end{Remark} 

We have considered $g_{11}(\theta)=f''(\theta)>0$ as a Hessian metric induced by the potential function $f(\theta)$.
However, we can also consider $g_{11}$ as a {\em squared Hessian metric}: Namely,
$g_{11}(\theta)=(\sqrt{f''(\theta)})^2$. 
Notice that $h(x)=\int \int \sqrt{f''(u)}\du$ is strictly convex and hence $h''(\theta)=\sqrt{f''(\theta)}>0$ is a potential function inducing a Hessian metric.

The following section, shows that squared Hessian metrics induced by multivariate potential functions are Euclidean metrics in arbitrary dimension with Karcher means expressed in the primal coordinate system as multivariate quasi-arithmetic means coinciding with left-sided Bregman centroids.

\section{Quasi-arithmetic Karcher means of the Euclidean metric}\label{sec:KarcherQAM}

Let $(M,g)$ be a $m$-dimensional Riemannian manifold equipped with a Hessian metric~\cite{Shima-2007} $g$ expressed in a global coordinate system $\theta(\cdot)$ as
$$
G(\theta)=[g_{ij}(\theta)]=\nabla^2 F(\theta), 
$$
where $F$ is a strictly convex and differentiable potential function of Legendre type~\cite{LegendreType-1967}.
We have $g_{ij}(\theta)=g(\partial_i,\partial_j)$ where $\partial_l=\frac{\partial}{\partial\theta_l}$.
The convex conjugate $F^*(\eta)=\sum_{i=1}^m \theta_i(\eta)\eta_i-F(\theta(\eta))$  is of Legendre type with $\eta(\theta)=\nabla F(\theta)$ and $\theta(\eta)=\nabla F^*(\eta)=(\nabla F)^{-1}(\eta)$. The Crouzeix identity~\cite{crouzeix1977relationship} is $\nabla ^2F(\theta)\nabla^2 F^*(\eta)=I_{m,m}$, where $I_{m,m}$ is the matrix identity of dimension $m$.

In general, the metric $g$ can be expressed in any other coordinate system, say $\xi(\theta)$, by using the covariant transformation on matrix $G(\theta)$:
$$
G_\xi(\xi)=J_\xi(\theta)^\top\, G(\theta(\xi))\, J_\xi(\theta),
$$
where $J_\xi(\theta)=\left[\frac{\partial\theta_i}{\partial\xi_j}\right]$ is the Jacobian matrix of $\theta(\xi)$.

For example, let $\xi=\eta$ be the dual parameterization of $\eta$.
We have $J_\eta(\theta)=\nabla_\eta \theta(\eta)=\nabla_\eta \nabla_\eta F^*(\theta)=\nabla^2_\eta F^*(\eta)$, and we get
$$
G_\eta(\eta)= \nabla^2_\eta F^*(\eta)^\top\, \nabla^2 F(\theta(\eta)) \nabla^2_\eta F^*(\eta) = \nabla^2_\eta F^*(\eta),
$$
since the Crouzeix identity holds: $\nabla^2_\eta F^*(\eta)^\top\, \nabla^2 F(\theta(\eta))=I_{m,m}$.

Now, consider squared Hessian metrics $g_\sqr$ defined as follows
$$
G_\sqr(\theta)=[g_\sqr(\partial_i,\partial_j)] = G(\theta)^2 = (\nabla^2 F(\theta))^2,
$$
and express $g_\sqr$ in the $\eta$-coordinate system:
$$
G_\sqr(\eta)= \nabla^2_\eta F^*(\eta)^\top\,  (\nabla^2 F(\theta(\eta)))^2   \nabla^2_\eta F^*(\eta) = I_{m,m}.
$$

Thus $g_\sqr$ is the Euclidean metric with Riemannian geodesic distance the Euclidean distance:
$$
\rho_{g_\sqr}(P_1,P_2)=\|\eta(P_1) -\eta(P_2) \|_2, \forall P_1,P_2\in M.
$$ 

The Euclidean distance can also be expressed equivalently in the primal $\theta$-coordinate system as:
$$
\rho_{g_\sqr}(P_1,P_2)=\| \eta(P_1) -\eta(P_2) \|_2= \|\nabla F(\theta_1) -\nabla F(\theta_2) \|_2,
$$
where $\theta_i=\theta(P_i)$.

\begin{prp}\label{prp:sqrHessianeucldist}
The Riemannian distance between $P_1$ and $P_2$ of a squared Hessian metric $(M,g_\sqr)$ induced by the potential function $F(\theta)$ is the Euclidean distance, expressed in the dual coordinate systems $\theta(\eta)=\nabla F^*(\eta)$ and $\eta(\theta)=\nabla F(\theta)$ as:
$$
\rho_{g_\sqr}(P_1,P_2)=\|\eta_1-\eta_2\|_2=\|\nabla F(\theta_1)-\nabla F(\theta_2)\|_2,
$$
where $\eta_i=\nabla F(\theta(P_i))$ and $\theta_i=\nabla F^*(\eta(P_i))$.
\end{prp}

It follows that the Riemannian center of mass (also called the Karcher mean~\cite{karcher2014riemannian}) $C$ of $n$ points $P_1,\ldots, P_n$ on $(M,g_\sqr)$ with $\theta$-coordinates $\theta_1=\theta(P_1),\ldots,\theta_n=\theta(P_n)$ and dual eta-coordinates  $\eta_1=\eta(P_1),\ldots,\eta_n=\eta(P_n)$:
$$
C=\arg \min_{P\in M} \frac{1}{n}\sum_{i=1}^n \rho_{g_\sqr}^2(P_i,P)
$$
is unique and expressed in the dual $\eta$-coordinate system as $\eta(C)= \frac{1}{n}\sum_{i=1}^n  \eta(P_i)= \frac{1}{n}\sum_{i=1}^n  \eta_i$, and in the primal 
 $\theta$-coordinate system as a multivariate quasi-arithmetic mean:
\begin{eqnarray}
\theta(C) &=& \nabla F^* \left( \frac{1}{n}\sum_{i=1}^n  \nabla F(\theta_i)\right),\\
&=& (\nabla F^{-1}) \left( \frac{1}{n}\sum_{i=1}^n  \nabla F(\theta_i)\right).
\end{eqnarray}

\begin{prp}\label{prp:mvqam}
The center of mass of  $n$ points $P_1,\ldots, P_n$ (with $\theta_i=\theta(P_i)$) on a squared Hessian manifold $(M,g_\sqr)$ with $G_\sqr(\theta)=(\nabla^2 F(\theta))^2$ for a strictly convex and differentiable potential function $F(\theta)$ is expressed as a quasi-arithmetic mean for the gradient $\nabla F(\theta)$:
$$
\theta(C) = (\nabla F^{-1}) \left( \frac{1}{n}\sum_{i=1}^n  \nabla F(\theta_i)\right).
$$
\end{prp}

Notice that in general, multivariate functions may not have global inverse functions (see the implicit function theorem~\cite{krantz2002implicit}). 
However, in the case of a Legendre-type convex function $F(\theta)$, the gradient map $\nabla F$ admits a global inverse $(\nabla F)^{-1}=\nabla F^*$ where $F^*$ denotes the convex conjugate.

Proposition~\ref{prp:mvqam} shows that the center of mass of squared Hessian metrics coincide with left-sided Bregman centroid~\cite{nielsen2009sided} induced by the potential function.

Notice that the Riemannian Euclidean geodesic in the $\eta$-coordinate system (i.e., Cartesian coordinate system) between $P_1$ and $P_2$ is
$$
\gamma_\eta(\eta_1,\eta_2;t)=(1-t)\eta_1+t\eta_2, 
$$
and the Riemannian Euclidean geodesic in the $\theta$-coordinate system is:
$$
\gamma_\theta(\theta_1,\theta_2;s)=\nabla F^{-1}(\nabla (1-s)F(\theta_1)+s\nabla F(\theta_2)),
$$
a weighted quasi-arithmetic mean.

Similarly, we may consider any other problem on the squared Hessian manifolds
 as Euclidean problems in the Cartesian coordinate system $\eta$ (e.g., the Fermat-Weber points~\cite{fekete2005continuous} or the Voronoi diagrams~\cite{toth2017handbook}).

Notice that the line elements expressed in the dual coordinate systems match:
$$
\ds_{g_\sqr}^2(\theta)=\dtheta^\top G_\sqr(\theta) \dtheta = \deta^\top G_\sqr(\eta) \deta =\ds_{g_\sqr}^2(\eta).
$$
However, it is different than the square of the line element of the Hessian metric $g$:
$\ds_{g_\sqr}\not= \ds_{g}$ when $F(\theta)$ is not a quadratic function.

Notice that as soon as the dimension $m>1$, a squared Hessian metric may not necessarily be a Hessian metric.

Consider the symmetrized Bregman divergence defined by
$$
S_F(\theta_1;\theta_2):=B_F(\theta_1:\theta_2)+B_F(\theta_2:\theta_1)=(\theta_2-\theta_1)^\top (\eta_2-\eta_1)=S_{F^*}(\eta_1;\eta_2).
$$

\begin{prp}[Theorem 3.2 of~\cite{IG-2016}]
The symmetrized Bregman divergence $S_F(\theta_1;\theta_2)$ can be interpreted as the energy induced by the Hessian metric $\nabla^2 F(\theta)$ on the primal/dual geodesics:
$$
S_F(\theta_1;\theta_2)=\int_0^1 \ds^2(\gamma(t))\dt=\int_0^1 \ds^2(\gamma^*(t))\dt.
$$ 
\end{prp}

Since the proof is omitted in~\cite{IG-2016}, we give a proof in~\ref{appendix:sbd} for sake of completeness.

\section{Connection with Fisher-Rao information geometry}\label{sec:IG}

In information geometry~\cite{IG-2016,nielsen2022many}, a statistical model $\{p_\lambda(x) \ :\ \lambda\in\Lambda\subset\bbR^D\}$ is called regular when its Fisher information matrix (FIM) $I(\lambda)=\mathrm{Cov}[\nabla_\lambda\log p_\lambda(x)]$ is positive-definite and can be expressed as $I(\lambda)=-E_\lambda[\nabla^2 \log p_\lambda]$.
For example, exponential families have densities (with respect to a base measure $\mu$ usually taken as the Lebesgue or counting measure) expressed canonically as 
$p_\lambda(x)=\exp(\inner{t(x)}{\theta(\lambda)}-F(\theta)+k(x))\, \dmu$ where $\theta(\lambda)$ denotes the natural parameter, $t(x)$ the sufficient statistic vector,
$F(\theta)=\log \int \exp(\inner{t(x)}{\theta(\lambda)}) \dmu(x)$ the log-normalizer also called cumulant function, and $k(x)$ is an auxiliary carrier term. 
Exponential families are regular models with Fisher information matrices expressed in the natural parameterization as $I(\theta)=\nabla^2 F(\theta)$.
It follows that the Fisher-Rao manifold induced by the FIM is a Hessian manifold~\cite{Shima-2007} (i.e., Fisher Riemannian metric is Hessian).
When $t(x)=x$, $\theta(\lambda)=\lambda$ and $k(x)=0$, the exponential family is called a natural exponential family.
Mixture families~\cite{IG-2016} and biparametric statistical models~\cite{amari2014curvature}  are  regular statistical models which also yield Hessian Fisher metrics. 

On a Hessian manifold, we have two dual affine coordinate systems $\theta(\eta)=\nabla F^*(\eta)$ and $\eta(\theta)=\nabla F(\theta)$ related by the Legendre-Fenchel transform. On exponential family Hessian manifolds, the dual parameterization $\eta$ can be interpreted as the expectation parameter: 
$\eta(\theta)=E_\theta[t(x)]$.

For uni-order exponential family models $\{p_\lambda\}$ ($D=1$), Proposition~\ref{prop:qam} in \S\ref{sec:dualscales} reports the Karcher midpoint as a quasi-arithmetic mean for the generator 
$h(\theta)=\int^\theta \sqrt{f''(u)} \du$, and Proposition~\ref{dualscales} further gives an interpretation of this midpoint using a dual quasi-arithmetic mean induced by the generator $h^\diamond(\eta)=\int^\eta \sqrt{{f^*}''} \du$  where $f^*$ is the  convex conjugate of $f$. 
Namely, these dual quasi-arithmetic means correspond to the Karcher mean on the Fisher-Rao manifold expressed in the dual canonical parameterizations.
The Fisher-Rao distance can be expressed equivalently as
$$
\rho(\lambda_1,\lambda_2)=|h(\theta(\lambda_1))-h(\theta(\lambda_2))|=|h^\diamond(\eta(\lambda_1))-h^\diamond(\eta(\lambda_2))|.
$$
Thus $h(\theta)$ and $h^\diamond(\eta)$ may be termed the dual Fisher-Rao coordinates of the exponential family.
Table~\ref{tab:ig} summarizes the Fisher information metric, Riemannian distance, and Riemannian Karcher midpoint in the dual parameterizations for a 1D Hessian manifold.

\begin{table}
\centering
\caption{Riemannian distance and Karcher mean in the 1D Hessian geometry induced by the pair of convex conjugate functions $f(\theta)$ and $f^*(\eta)$ 
with $h(\theta)=\int^\theta \sqrt{f''(u)} \du$ and $h^\diamond(\eta)=\int^\eta \sqrt{{{f^*}''}(u)} \du$. \label{tab:ig}}
\begin{tabular}{l|l|l}
 & $\theta={f^*}'(\eta)$ & $\eta=f'(\theta)$\\ \hline
Metric & $f''(\theta)$ & ${f^*}''(\eta)$\\
Riemann. dist. & $|h(\theta_1)-h(\theta_2)|$ & $|h^\diamond(\eta_1)-h^\diamond(\eta_2)|$\\
Karcher mean & $m_h(\theta_1,\theta_2)=h^{-1}\left(\frac{h(\theta_1)+h(\theta_2)}{2}\right)$ & 
$m_{h^\diamond}(\eta_1,\eta_2)=({h^\diamond})^{-1}\left(\frac{h^\diamond(\eta_1)+h^\diamond(\eta_2)}{2}\right)$\\ \hline
\end{tabular}

\end{table}

Let us reinterpret the two examples reported in \S\ref{sec:dualscales} from the lens of information geometry:

\begin{itemize}
\item Consider the Poisson family with probability mass functions expressed as  $\frac{\lambda^x}{x!} e^{-\lambda}$ with discrete support $\mathbb{N}\cup\{0\}$, where $\lambda>0$ is the intensity parameter.
The Poisson family is an exponential family with sufficient statistic $t(x)=x$, natural parameter $\theta=\log\lambda$, cumulant function $F(\theta)=e^\theta$, 
Fisher information $I(\theta)=e^\theta$, convex conjugate $F^*(\eta)=\eta\log\eta-\eta$ for $\eta=e^\theta$.
The base measure is the counting measure and the auxiliary carrier term is $k(x)=-\log x!$.
The Fisher-Rao distance~\cite{miyamoto2024closed} between two Poisson distributions of parameters $\lambda_1$ and $\lambda_2$ is 
$$
\rho(\lambda_1,\lambda_2)=2\, |\sqrt{\lambda_1}-\sqrt{\lambda_2}|.
$$
Example~\ref{ex1} reported that  $h(\theta)=2\exp(\frac{\theta}{2})$ and $h^\diamond(\eta)=2\sqrt{\eta}$ which yield closed-form formula for the Fisher-Rao centroid in the dual parameterizations. In the $\lambda$-parameterization, we get the Fisher-Rao midpoint $\lambda_{12}$ between $\lambda_1$ and $\lambda_2$ as:
$$
\lambda_{12}=\frac{(\sqrt{\lambda_1}+\sqrt{\lambda_2})^2}{4}.
$$

\item Consider the Bernoulli family with probability mass functions expressed as  $p^x(1-p)^{1-x}$ with discrete support $\{0,1\}$ for parameters $p$ ranging in $(0,1)$. The Bernoulli family is an exponential family with natural parameter $\theta(p)=\log \frac{p}{1-p}\in\bbR$ (and reciprocal function $p(\theta)=\frac{e^\theta}{1+e^\theta}$), cumulant function $F(\theta)=\log(1+e^\theta)$ and Fisher information $I(\theta)=\frac{e^\theta}{(1+e^\theta)^2}$ (or $I(p)=\frac{1}{p(1-p)}$).
The convex conjugate is $F^*(\eta)=\eta\log\eta+(1-\eta)\log(1-\eta)$ (Shannon negentropy).
The Fisher-Rao distance~\cite{miyamoto2024closed} between two Bernoulli distributions of parameters $p_1$ and $p_2$ is 
$$
\rho(p_1,p_2)=2\, |\arcsin(\sqrt{p_1})-\arcsin(\sqrt{p_2})|.
$$
Notice that $2\arcsin(\sqrt{p})$ is the classic Fisher's angular transformation~\cite{fisher1923xxi}.
Example~\ref{ex2} reported that  $h(\theta)=2\,\arctan(\exp(\theta/2))$ and $h^\diamond(\eta)=2\arcsin(\sqrt{\eta})$ which yield closed-form formula for the Fisher-Rao centroid in the dual parameterizations.
 In the $p$-parameterization, we get the Fisher-Rao midpoint $p_{12}$ between $p_1$ and $p_2$ as:
$$
p_{12}=\left(\sin\left(\frac{\arcsin(\sqrt{p_1})+\arcsin(\sqrt{p_2}) }{2}\right)\right)^2.
$$

\end{itemize}

Now, when the order $D$ of the exponential family is greater than one (i.e., multiparametric statistical model), 
 it follows from $\eta=\nabla F(\theta)$ that $\deta=\nabla^2 F(\theta)\dtheta$ and therefore we get
$$
\|\deta\|^2=\inner{\nabla^2 F(\theta)\dtheta}{\nabla^2 F(\theta)\dtheta}=\dtheta^\top (\nabla^2 F(\theta))^2 \dtheta =
\dtheta^\top G_{\mathrm{sqr}}(\theta) \dtheta.
$$

Thus the squared Hessian metric is the pullback of the Euclidean metric in the dual expectation coordinates $\eta$.

\vskip 0.3cm
Code snippets in the computer algebra system {\sc Maxima} (\url{https://maxima.sourceforge.io/}) demonstrating results of this paper are available at
\url{https://franknielsen.github.io/DualQAM}

\vskip 0.5cm
\noindent {Acknowledgements}. I am indebted to two anonymous reviewers for their  constructive comments and suggestions that improved this paper.

\bibliographystyle{plain}
\bibliography{ScaleFrechetBIB}

\appendix

\section{Symmetrized Bregman divergence}\label{appendix:sbd}

Consider the symmetrized Bregman divergence   defined by
$$
S_F(\theta_1;\theta_2):=B_F(\theta_1:\theta_2)+B_F(\theta_2:\theta_1)=(\theta_2-\theta_1)^\top (\eta_2-\eta_1)=S_{F^*}(\eta_1;\eta_2),
$$

\begin{prp}[Theorem 3.2 of~\cite{IG-2016}]
The symmetrized Bregman divergence $S_F(\theta_1;\theta_2)$ is interpreted as the energy induced by the Hessian metric $\nabla^2 F(\theta)$ on the primal/dual geodesics:
$$
S_F(\theta_1;\theta_2)=\int_0^1 \ds^2(\gamma(t))\dt=\int_0^1 \ds^2(\gamma^*(t))\dt.
$$ 
\end{prp}

\begin{proof}
The proof is based on the first-order and second-order directional derivatives.
The first-order directional derivative $\nabla_u F(\theta)$ with respect to vector $u$ is defined by 
$$
\nabla_u F(\theta)=\lim_{t\rightarrow 0} \frac{F(\theta+tv)-F(\theta)}{t}=v^\top \nabla F(\theta).
$$
 
The second-order directional derivatives $\nabla_{u,v}^2 F(\theta)$ is
\begin{eqnarray*}
\nabla_{u,v}^2 F(\theta) &=& \nabla_{u} \nabla_v F(\theta),\\
 &=& \lim_{t\rightarrow 0} \frac{v^\top \nabla F(\theta+tu)-v^\top\nabla F(\theta)}{t},\\
&=& u^\top \nabla^2 F(\theta) v.
\end{eqnarray*}

Now consider the squared length element $\ds^2(\gamma(t))$ on the primal geodesic $\gamma(t)$ expressed using the primal coordinate system $\theta$:
$\ds^2(\gamma(t))=\dtheta(t)^\top \nabla^2F(\theta(t)) \dtheta(t)$ with $\theta(\gamma(t))=\theta_1+t(\theta_2-\theta_1)$ and $\dtheta(t)=\theta_2-\theta_1$.
Let us express the $\ds^2(\gamma(t))$   using the second-order directional derivative:
$$
\ds^2(\gamma(t))=\nabla^2_{\theta_2-\theta_1}  F(\theta(t)).
$$
Thus we have $\int_0^1 \ds^2(\gamma(t))\dt=[\nabla_{\theta_2-\theta_1}  F(\theta(t))]_0^1$,
where the first-order directional derivative is $\nabla_{\theta_2-\theta_1}  F(\theta(t))=(\theta_2-\theta_1)^\top \nabla F(\theta(t))$.
Therefore we get $\int_0^1 \ds^2(\gamma(t))\dt=(\theta_2-\theta_1)^\top (\nabla F(\theta_2)-\nabla F(\theta_1))=S_F(\theta_1;\theta_2)$.

Similarly, we express the squared length element $\ds^2(\gamma^*(t))$ using the dual coordinate system $\eta$ as the second-order directional derivative of $F^*(\eta(t))$ with $\eta(\gamma^*(t))=\eta_1+t(\eta_2-\eta_1)$:
$$
\ds^2(\gamma^*(t))=\nabla^2_{\eta_2-\eta_1}  F^*(\eta(t)).
$$
Therefore, we have  $\int_0^1 \ds^2(\gamma^*(t))\dt=[\nabla_{\eta_2-\eta_1}  F^*(\eta(t))]_0^1=S_{F^*}(\eta_1;\eta2)$.
Since $S_{F^*}(\eta_1;\eta_2)=S_F(\theta_1;\theta_2)$, we conclude that
$$
S_F(\theta_1;\theta_2)=\int_0^1 \ds^2(\gamma(t))\dt=\int_0^1 \ds^2(\gamma^*(t))\dt
$$

In 1D, both pregeodesics $\gamma(t)$ and $\gamma^*(t)$ coincide. We have $\ds^2(t)=(\theta_2-\theta_1)^2 f''(\theta(t))=(\eta_2-\eta_1){f^*}''(\eta(t))$ so that we check that $S_F(\theta_1;\theta_2)=\int_0^1 \ds^2(\gamma(t))\dt=(\theta_2-\theta_1)[f'(\theta(t))]_0^1=(\eta_2-\eta_1)[{f^*}'(\eta(t))]_0^1=(\eta_2-\eta_1)(\theta_2-\theta_1)$.
\end{proof}

In Riemannian geometry, a curve $\gamma(t)$ parameterized by $t\in [0,1]$ minimizes the energy $E(\gamma)=\int_0^1 |\dot\gamma(t)|^2\dt$ if it minimizes the length $L(\gamma)=\int_0^1 \|\dot\gamma(t)\|\dt$ and $\|\dot\gamma(t)\|$ is constant. 
Using Cauchy-Schwartz inequality, we can show that $L(\gamma)^2\leq E(\gamma)$.

\section{Code snippets in the computer algebra system {\sc Maxima}}\label{sec:appendix}


\subsection{Exponential increasing scale of means}

The following code in the computer algebra system {\sc Maxima} (\url{https://maxima.sourceforge.io/}) demonstrates {\em experimentally} that the family of exponential quasi-arithmetic means form a scale:

\begin{verbatim}
/* quasi-arithmetic exponential means form an increasing scale of means */
kill(all);
fpprec:1000$
set_random_state(make_random_state(2025))$
a:-1+random (2.0); b:-1+random (2.0);
minalpha:-300$ maxalpha: 300$
exponentialMean(alpha,x,y):=(1.0/alpha)*log((exp(alpha*x)+exp(alpha*y))/2.0);
exponentialMean(minalpha,a,b); 
exponentialMean(maxalpha,a,b); 
\end{verbatim}

Running the above code yields the following output:

\begin{verbatim}
(%o0)	done
(a)	0.9369471273196543
(b)	-0.2288229220357811
(%o7)	exponentialMean(alpha,x,y):=1.0/alpha*log((exp(alpha*x)+exp(alpha*y))/2.0)
(%o8)	-0.2265124314339146
(%o9)	0.9346366367177878
\end{verbatim}

We check experimentally that for large negative values of $\alpha$, the exponential mean $m_{e_\alpha}$ tends to the minimum and 
for large positive values of $\alpha$, the exponential mean $m_{e_\alpha}$ tends to the maximum.
However, we observe experimentally that we need to take large values of $\alpha$ to approximate numerically the minimum and maximum values, and this requires multi-precision arithmetic.

\subsection{Radical decreasing scale of means}

The following code demonstrates experimentally that the radical means $\{m_{k_\alpha}\}_\alpha$ yields a decreasing scale of means:
\begin{verbatim}
/* Radical means generates a decreasing scale of means on the positive reals */
kill(all);
fpprec:30;
set_random_state(make_random_state(2025))$
a:random (1.0);
b:random (1.0);
f(alpha,x):=alpha**(1/x);
finv(alpha,x):=log(alpha)/log(x);
/* quasi-arithmetic means */
qam(alpha,x,y):=finv(alpha, (f(alpha, x)+f(alpha, y))/2);
qam(10**(-30),a,b)$ bfloat(%); 
qam(10**(30),a,b)$ bfloat(%); 
\end{verbatim}

Executing the above code yields the following output:

\begin{verbatim}
(%o0)	done
(fpprec)	30
(a)	0.9684735636598272
(b)	0.3855885389821094
(%o5)	f(alpha,x):=alpha^(1/x)
(%o6)	finv(alpha,x):=log(alpha)/log(x)
(%o7)	qam(alpha,x,y):=finv(alpha,(f(alpha,x)+f(alpha,y))/2)
(%o9)	9.59152532373302403747625205115b-1
(%o11)	3.87086223530841600144721384876b-1
\end{verbatim}
 
\subsection{Plotting scales of means}

The figure was obtained using the following code:

\begin{verbatim}
a:1; b:2;
radicalscale(beta) := beta/log( (exp(beta/a)+exp(beta/b))/2 );
exponentialscale(alpha):=(1/alpha)*log((exp(a*alpha)+exp(b*alpha))/2);
powerscale(alpha):=((a**alpha+b**alpha)/2)**(1/alpha);
plot2d([radicalscale(alpha),powerscale(alpha),exponentialscale(alpha)],[alpha,-20,20],
[legend, "radical mean", "power mean", "exponential mean"],
[xlabel, "t"], [ylabel, "midpoint c"],
[title, "Scale of quasi-arithmetic means"],[pdf_file, "scalemeans-20.pdf"]);
\end{verbatim}

\subsection{Fisher-Rao Karcher mean for the Bernoulli family}

\begin{verbatim}
 /* Bernoulli family */
kill(all)$
BernoulliFR(p1,p2):=2*abs(asin(sqrt(p1))-asin(sqrt(p2)))$
F(theta):=log(1+exp(theta))$
theta(p):=log(p/(1-p))$
eta(p):=p$
h(u):=2*atan(exp(u/2))$ hinv(u):=2*log(tan(u/2))$
hdiamond(u):=2*asin(sqrt(u))$ hdiamondinv(u):=(sin(u/2))**2$
BernoulliFRtheta(theta1,theta2):=abs(h(theta1)-h(theta2))$
BernoulliFReta(eta1,eta2):=abs(hdiamond(eta1)-hdiamond(eta2))$
p1:random(1.0)$p2:random(1.0)$
BernoulliFR(p1,p2);
BernoulliFRtheta(theta(p1),theta(p2));
BernoulliFReta(eta(p1),eta(p2));
FRcentroidTheta(t1,t2):=hinv((h(t1)+h(t2))/2)$
FRcentroidEta(e1,e2):=hdiamondinv((hdiamond(e1)+hdiamond(e2))/2)$
FRcentroidTheta(theta(p1),theta(p2))$ratsimp(%);
FRcentroidEta(eta(p1),eta(p2))$ratsimp(%);
p12:%$
BernoulliFR(p1,p12)$float(%);BernoulliFR(p12,p2)$float(%);
\end{verbatim}

 \end{document}